\documentstyle[aps,prl,floats,epsfig]{revtex}  
\def\DESepsf(#1 width #2){\epsfxsize=#2 \epsfbox{#1}}  

\begin{document}



\title{
 \quad\\[1cm] \Large
Search for Charmless Two-body Baryonic Decays of $B$ Mesons}

\author{The Belle Collaboration}

\maketitle


\begin{center}
  K.~Abe$^{7}$,               
  K.~Abe$^{40}$,              
  T.~Abe$^{41}$,              
  I.~Adachi$^{7}$,            
  Byoung~Sup~Ahn$^{15}$,      
  H.~Aihara$^{42}$,           
  M.~Akatsu$^{22}$,           
  Y.~Asano$^{47}$,            
  T.~Aso$^{46}$,              
  V.~Aulchenko$^{1}$
% ...
\end{center}

\small
\begin{center}
$^{1}${Budker Institute of Nuclear Physics, Novosibirsk}\\
% ...
$^{7}${High Energy Accelerator Research Organization (KEK), Tsukuba}\\
% ...
$^{15}${Korea University, Seoul}\\
% ...
$^{22}${Nagoya University, Nagoya}\\
% ...
$^{40}${Tohoku Gakuin University, Tagajo}\\
$^{41}${Tohoku University, Sendai}\\
$^{42}${University of Tokyo, Tokyo}\\
% ...
$^{46}${Toyama National College of Maritime Technology, Toyama}\\
$^{47}${University of Tsukuba, Tsukuba}\\
% ...
\end{center}

\normalsize

\tighten

\begin{abstract}

We report the results of a search for the rare baryonic decays 
$B^0 \to p\bar{p}$, $\Lambda\bar{\Lambda}$, 
and $B^+ \to p\bar{\Lambda}$.
The analysis is based on a
data set of $31.7\times 10^6 B\bar{B}$ events collected by the 
Belle detector at the KEKB $e^+e^-$
collider.  
No statistically significant signals are
found, 
and we set branching fraction upper limits
${\mathcal B}(B^0 \to p\bar{p}) < 1.2 \times 10^{-6}$, 
${\mathcal B}(B^0 \to \Lambda\bar{\Lambda}) < 1.0 \times 10^{-6}$, and 
${\mathcal B}(B^+ \to p\bar{\Lambda}) < 2.2 \times 10^{-6}$ 
at the 90\% confidence level.

\end{abstract}



%
%
%

\twocolumn[\hsize\textwidth\columnwidth\hsize\csname
@twocolumnfalse\endcsname      
\normalsize
\vskip2pc ]  

\pagestyle{plain}      


  The Belle collaboration recently reported the observation of the
  decay process $B^+ \to p\bar{p}K^+$\cite{ppk}, which is the first known
  example of a $B$ meson decay to a charmless final state
  containing baryons. In this paper we report the results of
  a search for the related two-body modes $B^0\to
  p\bar{p}$, $\Lambda\bar{\Lambda}$  and 
 $B^+\to p\bar{\Lambda}$\cite{foot}.  In the
  Standard Model, these decays are expected to proceed via
  color-suppressed  $b\to u$ tree diagrams (Figs.~1(a) and (c)) and
  $b\to s,~d$ penguin diagrams (Figs.~1(b) and (d)).  The search is based
  on a 29.4 fb$^{-1}$  sample of $e^+e^-$ data accumulated at the
  $\Upsilon(4S)$ resonance, which contains 31.7 million $B\bar{B}$
  pairs.   A previous search for these decays by the
  CLEO collaboration using a 5.41~fb$^{-1}$ sample of
  $\Upsilon(4S)$ data yielded 90\% confidence-level (C.L.) 
  upper limits\cite{CLEO}.


\begin{figure}[t]
\centering
\vspace{-1pc}
\epsfig{figure=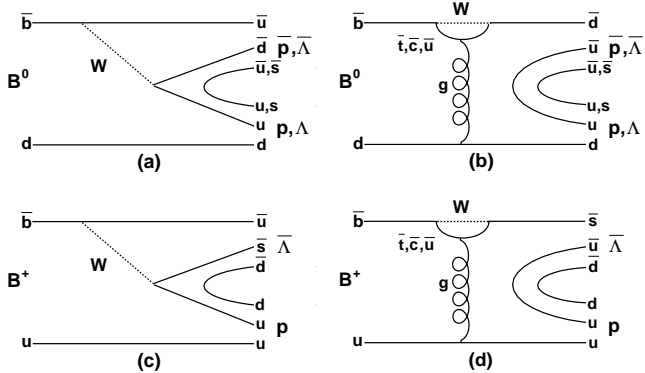,width=3.5in}
\centering
\caption{
Illustrative diagrams for $B$ decays to charmless baryon pairs.}
\label{Feyn}
\end{figure} 

Belle\cite{Belle} is a general purpose detector 
operating at the KEKB asymmetric $e^+e^-$ collider\cite{KEKB}. 
Tracking information is provided by a
silicon vertex detector  and a central drift chamber (CDC) in
a 1.5 Tesla magnetic field.
Hadron identification (PID) for $\pi/K/p$ 
discrimination is obtained from
CDC $dE/dx$ measurements, aerogel
${\check{\rm C}}$erenkov counter pulse heights, 
and timing information from the time-of-flight  system.
Electron identification is based mainly on CsI(Tl) electromagnetic
calorimeter and CDC $dE/dx$ information. $K_L$ and muons are 
identified by a 
system of resistive plate counters interleaved with 
the iron plates of the flux-return iron yoke.
 
The event selection criteria are based on tracking and PID
requirements, and are optimized using Monte Carlo (MC)
simulated event samples.

All primary charged tracks 
are required to satisfy the following track quality 
criteria based on
the track impact parameters relative to the   
interaction point (IP), which is determined run-by-run.
The $z$ axis is defined by the positron beam line.
The deviations from the IP position are required to be within
$\pm$0.05 cm in the transverse ($x$-$y$) plane, and within $\pm$2 cm
in the $z$ direction.
Tracks that satisfy the muon or electron identification
requirements are rejected.

Primary proton candidates are selected based
on $p/K/\pi$ likelihood functions
obtained from the hadron identification system. 
The selection criteria are
${L_p \over {L_p+L_K}}> 0.6 $ and  ${L_p \over {L_p+L_{\pi}}}> 0.6$,
where $L_{p/K/\pi}$ stands for the proton/kaon/pion likelihood. 

$\Lambda$ candidates are reconstructed via the $p\pi^-$ decay channel and
are selected using cuts on four parameters:
the angular difference between the $\Lambda$ flight direction
and the direction pointing from IP to the decay vertex in the
transverse plane; the distance between each track and the
IP in the transverse plane; the distance between the decay vertex and
the IP in the transverse plane; 
and the displacement in $z$ of the closest approach points of
the two tracks to the beam axis. The secondary protons are required
to have ${L_p \over {L_p+L_{\pi}}}> 0.6$. 
The $p\pi^-$ mass spectrum after the application of
the above selection criteria is shown
in Fig.~2 for a typical run period. 
The peak position is consistent with the nominal $\Lambda$ mass
\cite{pdg} and the mass resolution is about 0.9 MeV/$c^2$. 
Finally, we require the invariant mass of the $\Lambda$ candidate 
to be within $\pm$5 MeV/$c^2$ of the nominal
$\Lambda$ mass.


\begin{figure}[t]
\centering
\vspace{-1.5pc}
\epsfig{figure=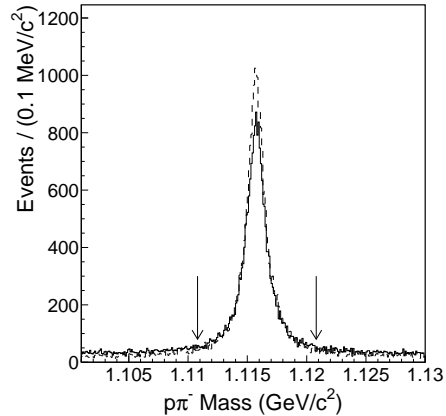,width=2.4in}
\label{Lam}
\centering
\caption{
The mass distribution of the selected $\Lambda \to p \pi^-$ 
candidates for 
a typical run period. The MC distribution is shown as a dashed
histogram.
The $\pm 5$ MeV mass window is indicated by the vertical arrows.}
\end{figure}

Due to the high momentum of the primary particles 
from these two-body decay modes,
the background from generic $B$ decays is negligible.
This is checked with  MC samples of $B^+B^-$ and  
$B^0\bar{B^0}$ pairs where the $B$-mesons decay dominantly via 
$b \to c$ processes. We also checked 
backgrounds using MC samples of charmless $B$ meson decays 
that include low multiplicity $B$ decays into 
final states with $\pi$,
$K$, $K^*$, $\rho$, $\omega$, $\phi$, $\eta$, and $\eta^\prime$ mesons.
Only the $p\bar{p}$ mode is found to have any background
contamination. However, the level of this contamination is negligible,
corresponding to less than one event for the current data set.

The main background is from continuum $q\bar{q}$ processes. 
This is confirmed using
an off-resonance data set (2.3 fb$^{-1}$ taken 60 MeV 
below the $\Upsilon(4S)$) and a MC sample of 
65 million continuum events.
These continuum events have a jet-like topology while
$B\bar{B}$ events are more spherical in the $\Upsilon(4S)$ center 
of mass (CM) frame. 
For continuum event rejection we use 
$\cos\theta_T$, the cosine of the angle between 
the direction of one primary decay particle and the 
thrust axis\cite{thrust} of the non-candidate tracks and showers.
This distribution is nearly flat for signal events and is strongly
peaked at $\pm$1 for continuum background.
We also use $\cos\theta_B$, the cosine of the angle between
the  $B$ candidate flight direction and
the positron beam direction. 
The signal has a $\sin^2\theta_B$ distribution
while the background is uniformly distributed.
We require the absolute values of $\cos\theta_T$ and $\cos\theta_B$ to be
less than 0.9 for $\Lambda\bar{\Lambda}$ decays and less than 0.8 for 
the other modes. For the latter case, the background reduction 
factor is more than five, while $\sim$70\% of the signal is retained.

\begin{table}[t]
\twocolumn[\hsize\textwidth\columnwidth\hsize\csname
@twocolumnfalse\endcsname    

\caption{Results of search for the exclusive baryon modes. 
The signal yields,
{\it Y}, and errors are determined from
maximum likelihood fits. The 90\% C.L. upper limits from the fits
and from the counting method are listed together. We quote the higher
values as our conservative estimates for upper limits. 
The efficiencies, $\varepsilon$, are obtained from MC.
The 90\% C.L. upper limits for the
 branching fractions, {$\mathcal B$},
determined by this experiment
are shown along with previous CLEO results.}
\medskip
\label{xsyield}
\begin{tabular}{cccccc}
{Mode} &
{\it Y} &
UL (fitting/counting) &              
{$\varepsilon$ (\%)} &
{$\mathcal B$}$(10^{-6})$ &
{CLEO $\mathcal B$}$(10^{-6})$
\\
\hline
$p\bar{p}$ &
$0.6^{+2.7}_{-0.6}$ &
7.0/9.7 &
$27.5 \pm 2.0$ &
$< 1.2$ &
$< 7.0$
\\
$\Lambda\bar{\Lambda}$ &
$0.0^{+0.5}_{-0.0}$ &
3.0/3.2 &
$10.8 \pm 1.1$ &
$< 1.0$ &
$< 3.9$
\\
$p\bar{\Lambda}$ &     
$1.0^{+2.5}_{-1.0}$ &
7.0/10.4 &
$16.2 \pm 1.4$ &
$< 2.2$ &
$< 2.6$ 
\end{tabular}
\vskip1pc
]       
\end{table}

We use the following two kinematic variables to identify the reconstructed
$B$ meson candidates: 
the beam constrained mass, $m_{\rm bc} =
\sqrt{E^2_{\rm beam}-p^2_B}$, and the energy difference, $\Delta{E} =
E_B - E_{\rm beam}$, where $E_{\rm beam}$, $p_B$ and $E_B$ are 
the beam energy, the
momentum and energy of the reconstructed $B$ meson in the $\Upsilon(4S)$
CM frame, respectively. 
We retain events with   5.20
GeV/$c^2$ $< m_{\rm bc} <$ 5.29 GeV/$c^2$ and 
$-0.2$ GeV $< \Delta{E} <$ 0.2 GeV.
The signal yield is extracted 
by maximizing the likelihood function 
$$ L = e^{-(s+b)}\prod_{i=1}^{N} [sP_s(m_{{\rm bc}_i},\Delta{E}_i)+
bP_b(m_{{\rm bc}_i},\Delta{E}_i)],$$
where $N$ is the total number of candidate events, $s(b)$ denotes 
the signal (background)
yield and $P_{s(b)}$ denotes the signal (background) probability
density function (pdf).


The $m_{\rm bc}$ and $\Delta{E}$ signal pdf's are determined by MC.
We use a Gaussian function as the signal
pdf for the $m_{\rm bc}$ distribution and a sum of two Gaussians 
for the  $\Delta{E}$ distribution.  The
Gaussian parameters (mean and $\sigma$) are
determined separately for each mode. 
Background shapes are determined from
events in sideband regions of $\Delta{E}$ and $m_{\rm bc}$ separately.
We adopt an empirical function\cite{Argus} to model 
the $m_{\rm bc}$ background shape 
(for events with  0.1 GeV $< |\Delta{E}| <$ 0.2 GeV)
and a first-order polynomial for the
$\Delta{E}$ background shape (for events with 5.20
GeV/$c^2$ $< m_{\rm bc} <$ 5.26 GeV/$c^2$).

Table~1 summarizes the results. The efficiencies for the
$\Lambda\bar{\Lambda}$ and $p\bar{\Lambda}$ modes include
the $\Lambda \to p\pi^-$ branching fraction (64\%). The efficiencies
are determined from signal MC 
with the identical event selection and fitting procedure as
for the data.
Figures~3 and 4 show the 
$m_{\rm bc}$ (with $|\Delta{E}| < 0.05$ GeV)  and $\Delta{E}$ (with
5.27 GeV/$c^2$ $< m_{\rm bc} < $ 5.29 GeV/$c^2$) projections for these 
three modes, respectively.
Projections of the fits are shown as smooth curves.
No statistically significant signals are found and we determine 
90\% C.L. upper
limits on the signal yields 
by integrating the likelihood function.
We also compute limits using a counting method\cite{Gary}. We define
a signal region by 5.27 GeV/$c^2$ $< m_{\rm bc} <$ 5.29 GeV/$c^2$ and
                $|\Delta{E}| < 0.05$ GeV, and treat the number of
background events from the maximum likelihood fit as a
                prediction for the background in this region. We then
                count the number of events actually observed, and apply
                the method of \cite{Gary} to obtain 90\% C.L. 
                intervals.     

To estimate the possible influence of fluctuations on the
determination of the upper limits,
we vary the parameters of the pdf's by one standard deviation and
change the form of the $\Delta{E}$ signal pdf to a single Gaussian.
The relative change of the upper limit is 25\%, which
is mainly due to the changes in the background shape. 
The upper limit determination  by the counting method is checked by
redefining the signal region (varying the $\Delta{E}$ range from 2$\sigma$ to 
4$\sigma$) and comparing the outcomes.  The
upper limits from both methods listed in column 3 of Table~1
include these fluctuations.
We quote the values from the counting method as the most conservative
upper limits.
Fit projections  with the signal yield fixed at the upper limit 
for each mode
are shown as the superimposed dashed curves in 
Figs.~3 and 4 for comparison.

\begin{figure}[t]
\centering
\vspace{-1.5pc}
\epsfig{figure=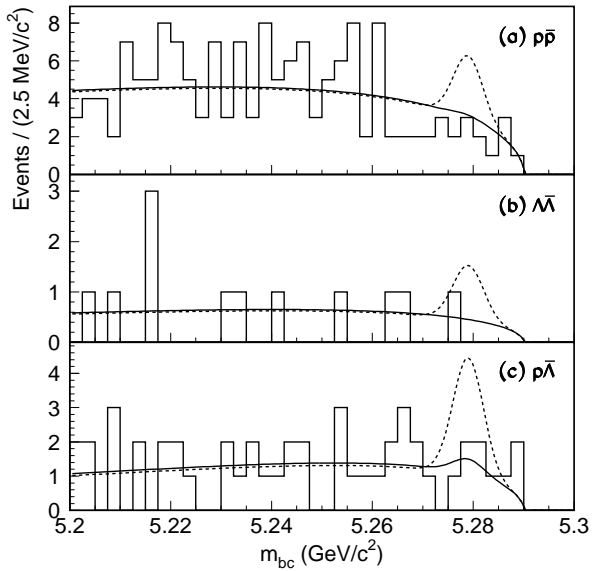,width=3.3in}
\label{mergemb}
\centering
\caption{
The distributions of $m_{\rm bc}$ for (a) $B^0\to p\bar{p}$, (b)
$B^0\to \Lambda\bar{\Lambda}$ and (c) $B^+\to p\bar{\Lambda}$ candidates.
         The solid curve is the projection of the maximum likelihood fit. 
The dashed curve shows the fit with the signal yield
fixed at the 90\% C.L. upper limit.
}
\end{figure}

\begin{figure}[t]
\centering
\vspace{-1.5pc}
\epsfig{figure=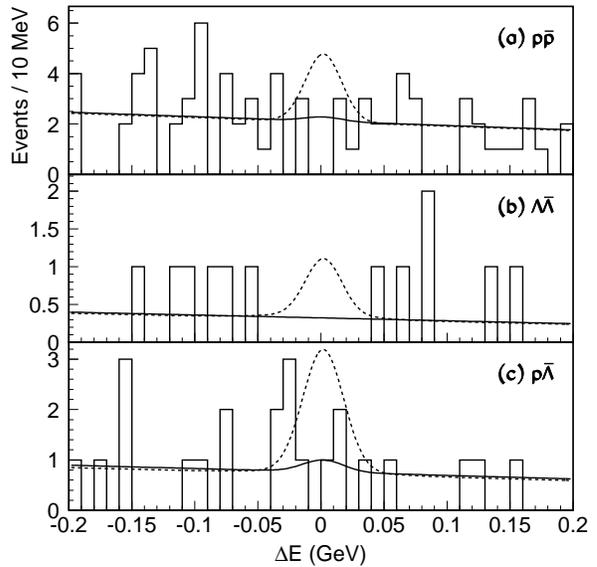,width=3.3in}
\label{mergede}
\centering
\caption{
The distributions of $\Delta{E}$ for (a) $B^0\to p\bar{p}$, (b)
$B^0\to \Lambda\bar{\Lambda}$ and (c) $B^+\to p\bar{\Lambda}$ candidates.
         The solid curve is the projection of the maximum likelihood fit.
The dashed curve shows the fit with the signal yield
fixed at the 90\% C.L. upper limit. 
}
\end{figure}

The systematic error due to the efficiency of the
proton identification ($p$-$K$ and $p$-$\pi$) criteria
is studied using $\Lambda$ samples. 
We vary the likelihood ratio requirement for protons and compare the
ratio of reconstructed $\Lambda$ yields in data and MC. The overall error 
is about 3\%. 
We include a 2\% error per track to account 
for the uncertainty in tracking efficiency. 
The $\Lambda$ reconstruction efficiency is checked by comparing the
flight distance distributions of data and MC. They agree very well 
and no additional error is assigned. 
The correlated parts of the errors are added
linearly to obtain
the overall uncertainty in the tracking efficiency and the uncertainty 
in the PID efficiencies (for $p$ and $\bar{p}$), then the
resulting errors are combined in quadrature. 
When determining the upper limit for the
branching fraction, the efficiency was reduced by one standard deviation.
The efficiencies and upper limits for all three decay modes
are listed in Table 1.


In summary, we have performed a search for 
the rare baryonic decays $B^0 \to p\bar{p}$, 
$\Lambda\bar{\Lambda}$, 
and $B^+ \to p\bar{\Lambda}$ with 
31.7 million $B\bar{B}$ events collected by the Belle detector 
at the KEKB $e^+e^-$
collider.  
No statistically significant signals are found for these modes, and we set
upper limits on their branching fractions at 
the 90\% C.L.. The upper limits are:
\begin{eqnarray*}
{\mathcal B}(B^0 &\to p\bar{p}&) < 1.2 \times 10^{-6};     \\
{\mathcal B}(B^0 &\to \Lambda\bar{\Lambda}&) < 1.0 \times 10^{-6}; \\
{\mathcal B}(B^+ &\to p\bar{\Lambda}&) <  2.2 \times 10^{-6}.
\end{eqnarray*}

These are currently the most stringent limits for these
decays. The limit 
on $B^0 \to p\bar{p}$ has been improved by a factor of six 
compared to the existing bound\cite{CLEO}. 
These limits are considerably below corresponding 
charmless mesonic branching fractions which are typically at the level
of $10^{-5}$.  
In contrast, the recent Belle  measurement of 
$B^+ \to p\bar{p}K^+$\cite{ppk} 
indicates a relatively large
branching fraction of $4.3 \times 10^{-6}$. 
The underlying physics\cite{Theory} is now being probed
with  rapidly accumulating data and new experimental results.



We wish to thank the KEKB accelerator group for the excellent
operation of the KEKB accelerator.
We acknowledge support from the Ministry of Education,
Culture, Sports, Science, and Technology of Japan
and the Japan Society for the Promotion of Science;
the Australian Research Council
and the Australian Department of Industry, Science and Resources;
the National Science Foundation of China under contract No.~10175071;
the Department of Science and Technology of India;
the BK21 program of the Ministry of Education of Korea
and the CHEP SRC program of the Korea Science and Engineering Foundation;
the Polish State Committee for Scientific Research
under contract No.~2P03B 17017;
the Ministry of Science and Technology of the Russian Federation;
the Ministry of Education, Science and Sport of Slovenia;
the National Science Council and the Ministry of Education of Taiwan;
and the U.S.\ Department of Energy.


\begin{thebibliography}{99}
\bibitem{foot} {
Charge-conjugate processes are included throughout the
                analysis.}
\end{thebibliography}
\end{document}